\documentstyle[aps,twocolumn]{revtex}

\begin{document}
\title {Non-perturbative approaches to magnetism in strongly
correlated electron systems}
\author{D. Vollhardt$^1$, N. Bl\"umer$^2$, K. Held$^1$, M. Kollar$^1$,
J. Schlipf$^1$, and M. Ulmke$^1$}
\address{$^1$ Theoretische Physik III, Elektronische Korrelationen und
Magnetismus, Universit\"at Augsburg, D-86135 Augsburg, Germany\\
$^2$ Department of Physics, University of Illinois at Urbana-Champaign,
1110 West Green Street, Urbana, IL 61801-3080, USA}
\maketitle

\begin{abstract}
  The microscopic basis for the stability of itinerant ferromagnetism
  in correlated electron systems is examined.  To this end several
  routes to ferromagnetism are explored, using both rigorous methods
  valid in arbitrary spatial dimensions, as well as Quantum Monte
  Carlo investigations in the limit of infinite dimensions (dynamical
  mean-field theory).  In particular we discuss the qualitative and
  quantitative importance of (i) the direct Heisenberg exchange
  coupling, (ii) band degeneracy plus Hund's rule coupling, and (iii)
  a high spectral density near the band edges caused by an appropriate
  lattice structure and/or kinetic energy of the electrons. We furnish
  evidence of the stability of itinerant ferromagnetism in the pure
  Hubbard model for appropriate lattices at electronic densities not
  too close to half-filling and large enough $U$. Already a weak
  direct exchange interaction, as well as band degeneracy, is found to
  reduce the critical value of $U$ above which ferromagnetism becomes
  stable considerably. Using similar numerical techniques the Hubbard
  model with an easy axis is studied to explain metamagnetism in
  strongly anisotropic antiferromagnets from
  a unifying microscopic point of view.\ \\ \\
  {71.27.+a,75.10.Lp}
\end{abstract}

\section{Introduction}

Even after several decades of research the microscopic foundations of 
itinerant ferromagnetism are not sufficiently understood. Indeed, in
contrast to other collective electronic phenomena such as 
antiferromagnetism or conventional 
superconductivity there exists a remarkable
gap between theory and experiment in this field. This has mainly to do
with the fact that itinerant ferromagnetism is a quantum-mechanical
{\it strong-coupling} phenomenon whose explanation requires 
the application of {\it non-perturbative} techniques. Thus it
belongs into the class of the most difficult many-body problems in
condensed-matter physics. Significant progress was made in this field
in the last few years due to the development and application of several new
analytic and numerical approaches. It is the purpose of this paper
to present and discuss some of these new, exciting results. 

In Sec.~2 we recapitulate the derivation of a general lattice
model for correlated electrons starting from a continuum model, and
discuss the various truncation steps which eventually lead to the Hubbard
model.
In particular, the implications of the Lieb-Mattis theorem on the impossibility
of ferromagnetism in these truncated models in $d=1$ dimension are analyzed. 
In Sec.~3 several microscopic mechanisms favoring ferromagnetism are 
discussed.
New and very recent results concerning ferromagnetism in Hubbard-type models
obtained by various non-perturbative methods are presented and put into 
perspective. Metamagnetic phase transitions
and the need for the application of non-perturbative techniques for these
investigations are the subject of Sec.~4. Finally, a conclusion is presented
in Sec.~5.

\section{Electronic Correlations and Magnetism}
\subsection{General lattice model}

Within the occupation number formalism the Hamiltonian for electrons with spin
$ \sigma $ interacting via a {\it spin-independent} interaction $ V^{ee} 
({\bf r} - {\bf r\, '})$ in the presence of an ionic lattice potential $
V^{ion} ({\bf r})$ has the form \cite{Lieb62,Hubbard63etc}
$\hat{H} = \hat{H_0} + \hat{H}_{\mbox{\scriptsize\em int}}$, where 
\begin{eqnarray}
\hat{H_0} &=& \sum_{\sigma} \int d^{3}r \hat{\psi}^{+}_{\sigma} ({\bf r}) 
\left[ - \frac{\hbar^2}{2m} \Delta + V^{ion} ({\bf r}) \right]
\hat{\psi}_\sigma ({\bf r} ) \label{h0}\\
\hat{H}_{\mbox{\scriptsize\em int}} &=& 
\frac{1}{2} \sum_{\sigma \sigma'} \int d^{3}r \int 
d^{3}r\, '   V^{ee} ({\bf r} - {\bf r\,'}) \hat{n}_{\sigma}
({\bf r} ) \hat{n}_{\sigma'} ({\bf r} \,')  \, . \label{hint}
\end{eqnarray}
Here $\hat{\psi}_{\sigma} ({\bf r} ) , \hat{\psi}^{+}_{\sigma} ({\bf r} )$ 
are the usual
field operators and $\hat{n}_{\sigma} ({\bf r}) = \hat{\psi}^{+}_{\sigma} ({\bf r} ) 
\hat{\psi}_{\sigma} ({\bf r} )$ is the local density. We note that the interaction 
term is diagonal in the space variables $ {\bf r}, {\bf r}\, ', $ i.e. it depends only
on the (operator-valued) densities of the electrons at site $ {\bf r}, {\bf r}\, '$
which interact via $V^{ee} ({\bf r} - {\bf r}\, '). $ The lattice potential
entering the non-interacting part (\ref{h0}) leads to a splitting of the
 parabolic dispersion into infinitely many
bands which we enumerate by the index $ \alpha. $ The non-interacting
problem is then characterized by the Bloch wave functions $ \phi_{\alpha 
{\bf k}} ({\bf r})$ and the band energies $ \epsilon_{\alpha {\bf k}}. $
We may introduce Wannier functions localized at site $ {\bf R_i} $ by
\begin{equation}
\chi_{\alpha i}({\bf r} ) = \frac{1}{\sqrt{L}} \sum_{{\bf k}} e^{-i 
{\bf k}\cdot{\bf R_i}} \ \phi_{\alpha{\bf k}}({\bf r})\, , \label{wann}
\end{equation}
where $L$ is the number of lattice sites, and thus construct creation and 
annihilation operators $\hat{c}^{+}_
{\alpha i \sigma}, \hat{c}^{\phantom +}_{\alpha i \sigma}$ for electrons with
spin $\sigma$ in the band $\alpha$ at site ${\bf R_i}$ as
\begin{eqnarray}
\hat{c}^{+}_{\alpha i \sigma} & = &\int d^{3}r \ \chi_{\alpha i} ({\bf r})\ 
\hat{\psi}^{+}_{\sigma} ({\bf r}) \nonumber \\  \longleftrightarrow 
\hat{\psi}^{+}_{\sigma} ({\bf r}) & = & \sum_{i\alpha}\ \chi^{*}_{\alpha i}
({\bf r})\hat{c}^{+}_{\alpha i \sigma} \, . \label{creation}
\end{eqnarray}
Thereby the Hamiltonian may be written in lattice representation as
\cite{Hubbard63etc}
\begin{eqnarray}
\hat{H} & = &\sum_{\alpha i j \sigma} t_{\alpha i j} \hat{c}^{+}_{\alpha i 
\sigma} \ \hat{c}^{\phantom +}_{\alpha j \sigma} \nonumber \\ +  
& & \frac{1}{2} \sum_{\alpha \beta \gamma \delta}
\sum_{i j m n} \sum_{\sigma \sigma'} {\cal V}^{\alpha\beta\gamma\delta}_{i
j m n} \ \hat{c}^{+}_{\alpha i \sigma} \ \hat{c}^{+}_{\beta j \sigma'} \
\hat{c}^{\phantom +}_{\delta n \sigma'} \ \hat{c}^{\phantom +}_{\gamma m 
\sigma} 
\, , \label{ham}
\end{eqnarray}
where the matrix elements are given by
\begin{eqnarray}
t_{\alpha i j} &=& \int d^{3}r \ \chi^{*}_{\alpha i} ({\bf r}) \left[ -
\frac{\hbar^2}{2m} \Delta + V^{ion} ({\bf r}) \right] \chi_{\alpha i} ({\bf r}) 
\label{matr1}\\
{\cal V}^{\alpha\beta\gamma\delta}_{i j m n} &=& \int d^{3}r \int d^{3}r\, ' 
V^{ee} ({\bf r} - {\bf r}\, ' ) \nonumber \\ & & \chi^{*}_{\alpha i} ({\bf r}) \ 
\chi^{*}_{\beta j} ({\bf r}\, ') \ \chi_{\delta n} ({\bf r}\, ') \ 
\chi_{\gamma m} ({\bf r}) \, . \label{matr2}
\end{eqnarray}
We note that in contrast to the field-operator representation defined in the
continuum, the Wannier representation does {\it not} lead to a site-diagonal
form of the electron-electron interaction, i.e. the interaction does not only
depend on the {\it densities} $\hat{n}_{i \sigma} = \hat{c}^+_{i \sigma} 
\hat{c}^{\phantom +}_{i \sigma}$ but contains explicit off-diagonal 
contributions which will be discussed later.  

\subsection{One-band models}

The Hamiltonian (\ref{ham}) is too general to be tractable in dimensions 
$d > 1$. Hence it has to be simplified using physically motivated truncations 
\cite{Hubbard63etc}.
In particular, if the Fermi surface lies within a single conduction band, 
and if this band is well separated from the other bands and the interaction 
is not too strong, it may be justified to restrict the discussion to a 
{\it single} band ($ \alpha = \beta = \gamma = \delta = 1$). 
In this case (\ref{ham}) reduces to
\begin {eqnarray}
\hat{H}_{\mbox{\scriptsize\em 1-band}} & = &
\sum_{ij\sigma} t_{ij} \hat{c}^+_{i\sigma} \hat{c}^{\phantom +}_{j\sigma} 
\nonumber \\
& & + \frac{1}{2} \sum_{ijmn} \sum_{\sigma \sigma'} {\cal V}_{ijmn} 
\hat{c}^+_{i\sigma} \hat{c}^+_{j\sigma'} \hat{c}^{\phantom +}_{n\sigma'} 
\hat{c}^{\phantom +}_{m\sigma} \, .
\label{1band}
\end {eqnarray}
For most purposes this single-band Hamiltonian is still too complicated.
Taking into account the weak overlap between neighboring orbitals in a 
tight-binding description one may expect that the overlap between 
nearest-neighbors
is most important. Hence the site-indices in (\ref{1band}) are restricted 
to nearest-neighbor positions. In the interaction this leaves us with a purely
local contribution ${\cal V}_{iiii} = U$, the Hubbard term, and the four
nearest-neighbor contributions ${\cal V}_{ijij} = V$, ${\cal V}_{iiij} = X$,
${\cal V}_{ijji} = F$, and ${\cal V}_{iijj} = F'$, which are 
{\it off}-diagonal in the site indices. The remaining one-band, 
nearest-neighbor Hamiltonian has the form 
\cite{Hubbard63etc,Caron68,Kivelson87etc,Gammel88etc,Baeriswyl88,Hirsch89etc,Painelli89,Strack94}
\begin {equation}
\hat{H}^{\mbox{\scriptsize\em NN}}_{\mbox{\scriptsize\em 1-band}} = 
\hat{H}_{\mbox{\scriptsize\em Hub}} + 
\hat{V}^{\mbox{\scriptsize\em NN}}_{\mbox{\scriptsize\em 1-band}} 
\label{nn}
\end {equation}
where
\begin {equation}
\hat{H}_{\mbox{\scriptsize\em Hub}} = -t \sum_{\langle i,j\rangle ,\sigma} 
(\hat{c}^+_{i\sigma} \hat{c}^{\phantom +}_{j\sigma} + {\rm h.c.})
+ U \sum_{i} \hat{n}_{i\uparrow} \hat{n}_{i\downarrow} \label{hub}
\end {equation}
is the Hubbard model and 
\begin {eqnarray}
\hat{V}^{\mbox{\scriptsize\em NN}}_{\mbox{\scriptsize\em 1-band}} & = & 
\sum_{\langle i,j\rangle} \ \lbrack 
V \hat{n}_i \hat {n}_j \nonumber\\
 & & + X \sum_{\sigma} (\hat{c}^+_{i\sigma} 
\hat{c}^{\phantom +}_{j\sigma} + {\rm h.c.}) 
(\hat {n}_{i\, -\sigma}+\hat {n}_{j \, -\sigma}) \nonumber\\
& & + F{'}  (\hat{c}^+_{i\uparrow} \hat{c}^+_{i\downarrow}
\hat{c}^{\phantom +}_{j\downarrow} 
\hat{c}^{\phantom +}_{j\uparrow}+ {\rm h.c.})\nonumber\\
& &  -2 F ({\bf {\hat{S}}}_{i} {\bf {\hat{S}}}_{j} 
+\frac {1}{4}\hat {n}_{i}\hat {n}_{j})\rbrack \label{vnn} 
\end {eqnarray}
is the contribution of nearest-neighbor interactions. Here
$\hat{n}_{i} = \sum_{\sigma} \hat{n}_{i\sigma}$ 
and ${\bf {\hat{S}}}_{i}=1/2 \sum_{\sigma\sigma'} \hat{c}^+_{i\sigma}
\tau^{\phantom +}_{\sigma\sigma'} \hat{c}^{\phantom +}_{i\sigma'}$,
where $\tau$ denotes the vector of Pauli matrices.
In (\ref{vnn}) the $V$-term describes a density-density 
interaction, the $X$-term is a bond-charge--site-charge interaction 
(``density-dependent hopping''), the $F'$-term describes the hopping of 
local pairs consisting of an up and a down electron and, finally, 
the $F$-term corresponds to the
{\it direct} Heisenberg exchange which is generically ferromagnetic in nature.
The occupation number formalism illustrates particularly clearly that a
spin-independent interaction plus the Pauli principle is able to lead to a 
mutual orientation of spins. Of all interactions in {(\ref{nn})} the
Hubbard interaction $U$ is certainly the strongest. 
Hence, in a final truncation
step one may try to neglect even the nearest-neighbor interactions and 
retain only the on-site interaction $U$. This leaves us with the Hubbard
model {(\ref{hub})}, the simplest correlation model for lattice electrons
\cite{Hubbard63etc,Gutzwiller63etc,Kanamori63}.

\subsubsection{The Hubbard model}

The Hubbard model was originally introduced in an attempt to understand
itinerant ferromagnetism in $3d$-transition metals
\cite{Hubbard63etc,Gutzwiller63etc,Kanamori63}.
The expectation was
that in this model ferromagnetism would arise naturally since in a polarized
state the electrons do not interact at all. However, it soon became clear
that in a ferromagnetic state the {\it{kinetic}} energy is also reduced.
This makes the stability of ferromagnetism in the Hubbard model a particularly 
delicate problem. Indeed, the kinetic energy with nearest-neighbor hopping 
usually favors {\it{anti}}\-ferro\-magnetism. At half-filling $(n = 1)$
and on bipartite
lattices antiferromagnetism is a generic effect since it appears both
at weak coupling (Hartree-Fock or Slater mean-field theory) and strong
coupling (Anderson's ``superexchange'' mechanism). Hence it arises 
naturally in any perturbational approach and, in particular, is tract\-able
by renormalization group methods \cite{Shankar94}.
By contrast, ferromagnetism is a non-trivial
strong-coupling phenomenon which cannot be investigated by any stand\-ard
perturbation theory.

The above discussion shows that, to understand the microscopic origin
of itinerant ferromagnetism, non-per\-tur\-bative techniques are required.
Unfortunately, there are not many approaches of this type available;
rigorous mathematical methods 
(for recent reviews see ref.~\cite{Mielke93,Lieb95,Strack95}), 
large-scale numerical methods
\cite{Gammel88etc,Macedo92,Hirsch94etc,Hlubina96etc,Daul97etc},
and variational approaches 
\cite{Shastry90,Fazekas90,MuellerHartmann93,Oles84,Fulde95,Buenemann96etc} 
are the ones most frequently used.

As to rigorous results about ferromagnetism in the Hubbard model
the Lieb-Mattis theorem \cite{Lieb62} of 1962 is one of the most famous.
It proves that for spin- and velocity-independent
forces between electrons ferromagnetism cannot occur in one spatial 
dimension. This theorem applies to the general Hamiltonian
(\ref{ham}) with infinitely many bands. On the other hand the general
{\it single-}band model (\ref{1band}) is {\it not} covered by the theorem
unless (a) the interaction matrix element ${\cal V}_{ijmn} $ is site-diagonal
such that the interaction depends only on densities ${\hat n}_{i}$, and
(b) the hopping and interaction does not extend beyond nearest neighbors.
Hence the theorem applies to the model (\ref{nn}) with $U , V \;{\ne}\; 0$ but
$X,F,F'=0$, and thereby also to the Hubbard model $(V = 0)$. We 
note that any critique of the single-band model (\ref{nn})
in view of the fact that it can lead to ferromagnetism in $d = 1$ in
contrast to the Lieb-Mattis theorem would apply even more so to
the Hubbard model since the latter is
a particularly special single-band model. In other words:
the fact that the Lieb-Mattis theorem 
applies to the Hubbard model but \em not \em to the \em general \em
single-band model (\ref{nn}) does not make the Hubbard model a 
``more physical'' model than (\ref{nn}); after all it is only a 
special case of (\ref{nn}).

Another well-known theorem, that by Nagaoka \cite{Nagaoka66} of 1966, 
provides explicit, albeit highly idealized, 
conditions under which ferromagnetism {\it is}
stable in the Hubbard model with nearest-neighbor hopping. It proves
that for $U =  \infty$ the microscopic degeneracy of the ground state at
half-filling (number of electrons $N$ = number of lattice sites $L$) is lifted
by a single hole, i.e. when one electron is removed $(N = L - 1)$. In this 
case a saturated ferromagnetic ground state is stable for any value of the 
hopping $t$ 
on simple cubic and bcc lattices, and for $t <  0$ on fcc and hcp lattices.
For the Nagaoka mechanism to work the lattice needs to contain loops along 
which the holes can move. Once the hole moves, the maximal overlap between 
the initial
and final state clearly occurs in a {\it ferromagnetic} configuration.
The problem is that Nagoaka's proof does not even extend to two holes,
that a single hole is thermodynamically irrelevant, and that the limit of
$U=\infty$ is highly unrealistic.

  \section{Microscopic Mechanisms Favoring Ferromagnetism}
  The Hubbard interaction is the result of an extreme truncation of
  the interaction in the general Hamiltonian (\ref{ham}). All
  interactions beyond the purely local part (e.g.\ nearest-neighbor
  density-density interactions, direct exchange, band degeneracy and
  the associated Hund's rule couplings) are totally neglected. The
  Hubbard interaction is therefore very unspecific --- it does not
  depend on the lattice at all and hence not on the spatial dimension.
  The lattice structure only enters via the kinetic energy. Therefore
  the stability of ferromagnetism in the Hubbard model can be expected
  to depend in a sensitive way on the precise form of the kinetic
  energy \cite{Mielke93,Lieb95,Nagaoka66,Tasaki95,Penc96,Fazekas96etc}.
  Strategies to find the essential ``kick'' for ferromagnetism in the
  Hubbard model and more general models should then proceed in
  different directions: one may (i) keep interactions beyond the
  Hubbard-$U$ (in particular the direct exchange term $F$ in
  (\ref{nn})), (ii) keep band degeneracy and Hund's rule couplings,
  or (iii) find an appropriate kinetic energy and lattice
  structure. We will now discuss several recent results obtained along
  these lines.

  \subsection{The importance of the Heisenberg exchange interaction}
  The Heisenberg exchange interaction, caused by direct
  quantum-mechanical exchange of electrons at nearest-neighbor
  positions (the $F$-Term in (\ref{vnn}) with $F>0$), favors the
  alignment of the electronic spins and hence supports
  ferromagnetism in a straightforward way
  \cite{Hirsch89etc,Strack94,Strack95}. However, since this
  interaction is rather weak (Hubbard \cite{Hubbard63etc} estimated
  $F\sim\frac{1}{40}$ eV for $3d$-metals, such that $F\ll U$)
  it cannot be the {\em sole} origin of itinerant ferromagnetism in
  systems like Fe, Co, Ni. Nevertheless it may be qualitatively
  important, since it may well give a correlated system with more or
  less strong ferromagnetic tendencies the ultimate push and trigger
  ferromagnetism in spite of its smallness. It is therefore
  unjustified to neglect the exchange interaction for merely {\em
    quantitative} reasons. This becomes particularly clear in the
  limit of large $U$ (with $U\gg|t|$, $|V|$, $|X|$, $F$, $F'$) close
  to $n=1$, when the one-band model (\ref{nn}) can be transformed
  into an effective $t$-$J$-model \cite{Gammel88etc,Strack94etc}
  \begin{eqnarray}\label{extendedtj}
    \hat{H}^{\mbox{\scriptsize\em NN}}
    _{\mbox{\scriptsize\em 1-band,$tJ$}}
    &=&-t\!\!\sum_{\langle i,j\rangle,\sigma}
          \!\!\hat{P}(\hat{c}^+_{i\sigma}
          \hat{c}^{\phantom +}_{j\sigma}
          +\mbox{h.c.})\hat{P}
    +J \sum_{\langle i,j\rangle}\hat{\bf S}_{i}\hat{\bf S}_{j}
  \end{eqnarray}
  where $\hat{P}$ projects onto the subspace without doubly occupied
  sites. The effective exchange coupling
  \begin{equation}\label{effectivej}
    J=\frac{4t^2}{U}\Big[\Big(1-\frac{X}{t}\Big)^2-\frac{FU}{2t^2}\Big]
  \end{equation}
  has an antiferromagnetic part, due to Anderson's
  superexchange but modified by the $X$-term,
  and a ferromagnetic part, due to the direct Heisenberg
  exchange. Hence for large enough Heisenberg exchange $F$ and/or
  Hubbard repulsion $U$ the exchange becomes effectively
  ferromagnetic. This effect is completely neglected in the Hubbard
  model where even in the Nagaoka-limit ($U=\infty$) the
  dimensionless parameter $FU/t^2$ is kept zero!

  \subsubsection{Generalization of Nagaoka's theorem} 
  If the Heisenberg exchange coupling is taken into account,
  it is possible to generalize Nagaoka's theorem to
  $U<\infty$ \cite{Kollar96}. We start with the
  Hamiltonian $\hat{H}^{\mbox{\scriptsize\em
      NN}}_{\mbox{\scriptsize\em 1-band}}$,
  (\ref{nn}).  This Hamiltonian, like the Hubbard
  Hamiltonian, commutes with the total spin $\hat{\bf
    S}=\sum_i\hat{\bf S}_i$.  The eigenvalues of
  $\hat{\bf S}^2$ are denoted by $S(S+1)$. We are
  concerned with saturated ferromagnetic states with one
  hole below half-filling, i.e.\ with largest possible
  eigenvalue $S_{\mbox{\scriptsize\em max}}\equiv
  N/2=(L-1)/2$.  There are $2S_{\mbox{\scriptsize\em
      max}}+1=L$ such states with the same energy
  eigenvalue.  It can be shown that the ground states of
  $\hat{H}_{\mbox{\scriptsize\em NN}}$ with one hole
  (i.~e.\ $N=L-1$) have maximum total spin
  $S=S_{\mbox{\scriptsize\em max}}=(L-1)/2$ and are
  non-degenerate (apart from the above-mentioned
  $(2S_{\mbox{\scriptsize\em max}}+1)$-fold spin
  degeneracy) in the following cases \cite{Kollar96}:
  
  Case 1:~On any lattice, if $F>0$, $t\leq0$ and
  (a)~$X\neq t$ and $U>U_c^{(1)}$, or (b)~$X=t$ and
  $U\geq U_c^{(2)}$.
  
  Case 2:~On lattices with loops, if $X=t<0$, $F=0$, and
  $U>U_c^{(2)}$.
  
  In both cases $t>0$ is allowed if the lattice is
  bipartite.
  
  These results are summarized in Table~\ref{crtab}.  The
  constants $U_c^{(1)}$ and $U_c^{(2)}$ are given by
  \begin{eqnarray}
    \label{bounds1}
    U_{c}^{(1)}&=&Z\Big(2|t|+\Big|V-F-2|t|\Big|
    \nonumber\\
    &+&{\frac{(X-t)^2}{F}}+
    \Big|F'-{\frac{(X-t)^2}{F}}\Big|\Big)\,,\\ 
    \label{bounds2}
    U_{c}^{(2)}&=&Z\Big(2|t|+
    \Big|V-{\frac{F}{2}}-2|t|\Big|+|F'|\Big)
  \end{eqnarray}
  \noindent where $Z$ is the number of nearest neighbors. 
  Hence, if $F>0$ ferromagnetic ground 
  states are stable on any lattice for $U$ larger than a
  {\em finite} critical value.  For $F\to0^+$ we have
  $U_c^{(1)}\to\infty$, thus yielding Nagaoka's condition
  for the pure Hubbard model.  This shows once more that
  the Heisenberg interaction $F$, which is neglected in
  the Hubbard model, provides an obvious mechanism for
  stabilizing ferromagnetic ground states at finite $U$.
  Note that since $X$ and $t$ are expected to be of the
  same order of magnitude, the sensitive dependence on
  $F$, due to the term $(X-t)^2/F$, may cancel from
  $U_c^{(1)}$. The dependence of $U_c$ on $t,V,F$ is
  depicted in Fig.~\ref{uvf}.  The case $X=t$ is special,
  since in this case the stability of ferromagnetism can
  be achieved either by $F>0$, or by $F\geq0$ and $t<0$
  if the lattice has loops.
  Note that the case $F>0$ is {\em not connected}
  to the case $F=0$ by a limiting procedure, since only in
  the latter case the lattice is required to have loops.

  The critical couplings $U_c^{(1)}$ and $U_c^{(2)}$ are
  sums of terms, each of which corresponds to a typical
  energy scale. This means that the on-site interaction
  $U$ has to be larger than the energy describing the
  paramagnetic state (bandwidth $\sim Z|t|$), as well as
  the threshold energies for the onset of a
  charge-density wave or phase separation ($\sim Z|V|$),
  $\eta$-pairing superconductivity \cite{deBoer95a} ($\sim
  Z|F'|$), and a spin-density wave ($\sim(X-t)^2/F$).
  However, these terms do not enter separately, but
  appear in combinations, i.~e.\ the effects interfere, as
  should be expected.
  
  The above conditions are {\em sufficient} conditions.
  The occurrence of ground states with maximum spin
  outside the above parameter region is not ruled out. As
  in Nagaoka's theorem for the pure Hubbard model, the
  ferromagnetic ground state is an itinerant state with
  non-zero kinetic energy, but the proof of its stability
  cannot yet be extended to doping beyond a single hole.

\subsubsection{Magnetic phase diagram within the dynamical mean-field theory}

Even for the simplest electronic correlation model, the Hubbard model,
exact solutions are not available in $d=2,3$ dimensions, and numerical
methods --- whether exact diagonalizations or Quantum Monte Carlo
(QMC) techniques --- are limited by the smallness of the systems that
can be studied. Hence one would like to construct, at least, a
thermodynamically consistent mean-field theory which is valid also at
strong coupling. Such a (non-perturbative) approximation is provided
by the exact solution of a model in $d=\infty$. It is now known that
in the limit $d\to\infty$~\cite{Metzner89etc} one obtains a {\em
  dynamical} mean-field theory for Hubbard-type models where the
spatial dependencies become local, but all quantum fluctuations of the
$d$-dimensional model are
included~\cite{MuellerHartmann89etc,Janis91,Georges92a,Jarrell92,Janis92a}.
In fact the problem is equivalent to an Anderson impurity model
complemented by a self-consistency
condition~\cite{Georges92a,Jarrell92}, leading to dynamical mean-field
equations which can be solved numerically, e.g.~within a
finite-temperature QMC scheme~\cite{Hirsch86}; for reviews see
refs.~\cite{Pruschke95,Georges96}. We employed this numerical approach
(for details see ref.~\cite{Ulmke95a}) to investigate the influence of
the direct exchange interaction $F$ on the stabilization of
ferromagnetism in the Hubbard model, using the one-band Hamiltonian
(\ref{nn})-(\ref{vnn}) with $X=F'=0$~\cite{Bluemer97etc}. To take the
limit $d\to\infty$ the couplings in (\ref{nn})-(\ref{vnn}) have to be
scaled appropriately~\cite{Metzner89etc,MuellerHartmann89etc}, i.e.
\begin{equation}
\label{scale}
t=\frac{t^*}{\sqrt{Z}},\ F=\frac{F^*}{Z},\ V=\frac{V^*}{Z}.
\end{equation}
In the following we set $t^*\equiv 1$. In the limit $d\to\infty$ the
$V^*$-term~\cite{MuellerHartmann89etc} and $F^*$-term reduce to their
Hartree-contributions. Hence their influence is that of a generalized,
i.e.~spin- and site-dependent, chemical potential. In the homogeneous
phase the spin- and site-dependent terms vanish. This implies that the
nearest-neighbor interactions $F^*$ and $V^*$ become important only in
the symmetry-{\em broken} phase. Consequently the susceptibilities of
the model (\ref{nn})-(\ref{vnn}) with $X=F'=0$ may be calculated from
the pure Hubbard model~\cite{Bluemer97etc}; this simplifies the matter
considerably.

The phase boundaries between the paramagnetic, antiferromagnetic and
ferromagnetic phases may then be calculated in principle as
follows~\cite{Bluemer97etc}: (i) QMC simulations are performed in the
homogeneous phase of the pure Hubbard model for given $U$, temperature
$T$, filling $n$, and number of Matsubara frequencies $\Lambda$; (ii)
for arbitrary values of $F^*$, $V^*$ the appropriate susceptibilities
are calculated; (iii) the inverse susceptibilities are extrapolated to
$\Lambda\to\infty$, if they are negative the homogeneous phase is
found to be unstable; (iv) to obtain ground state properties the
calculated quantities have to be extrapolated to $T\to 0$. The results
of these calculations for $n=1$, $T=0$ and a semi-elliptic DOS are
collected in Fig.~\ref{phaseT0}, where the exchange coupling $F^*$ is
plotted versus the Hubbard interaction $U$. We neglect the
density-density terms in (\ref{vnn}), since they become important
only for $V^*-F^*/2>U$~\cite{vanDongen91etc,Bluemer97etc}. The
solid line marks the phase boundary to the ferromagnetic phase. As
expected the critical value of $F^*$, $F^*_c$, decreases with $U$: it
depends on $U$ as $(F^*_c(0)-F^*_c(U))\propto U$ for small $U$
(Hartree-Fock limit) and as $F^*_c(U)\propto 1/U$ for large $U$
(Heisenberg limit).  At $U=12$ the value of $F^*$ necessary to induce
ferromagnetism is seen to be as small as $F^*_c\sim 0.1$. This shows
how important even a weak exchange coupling $F$ is for the stability
of ferromagnetism.  For $U>4$ there may well be a direct transition
between the antiferromagnetic and the ferromagnetic phases.

The shape of the DOS and the band filling $n$ are very important
factors concerning the stability of ferromagnetism, as will be discussed below
(Sec.~3.3). Indeed, a {\em symmetric} DOS and filling $n=1$, as used
in the above calculation, are especially disadvantageous for
ferromagnetism, partly because antiferromagnetism will be the generic
magnetic order in this case.

  \subsection{Band degeneracy and Hund's rule coupling}
  
  Another important route to ferromagnetism may be
  taken by considering more than one energy band,
  namely by starting from $M>1$ Wannier (or tight-binding)
  orbitals. It is known from atomic magnetism that
  there are ferromagnetic couplings between
  electrons on the same atom leading to Hund's
  rules. These on-site ``Hund's rule couplings''
  express the fact that by putting electrons in a maximum
  spin state an atomic exchange energy
  may be gained by the following mechanism. A spin
  wave function with maximum spin is always symmetric.
  Therefore, for the total wave function to be
  antisymmetric, the coordinate wave function must
  be antisymmetric.  This reduces the
  probability for electrons to come close to each
  other which in turn lowers the Coulomb energy
  between them.  This consideration establishes
  that in general there will be ferromagnetic
  on-site interactions {\em even in a bulk system.}
  Whether the presence of these terms is sufficient
  for ferromagnetism to appear in the ground state
  is, however, strongly dependent on their relative
  strength compared with the kinetic energy, on the
  lattice structure, electron density, etc.
  
  Let us therefore take a closer look at the
  terms in the Hamiltonian (\ref{ham}) in the
  case of $M$ relevant bands.  In this case we
  have to retain the band index
  $\alpha=1,\ldots M$.  Now there exist
  important on-site interactions even beyond
  the Hubbard interaction $U={\cal
    V}_{iiii}^{\alpha\alpha\alpha\alpha}$,
  namely the following couplings that are
  off-diagonal in the {\em band} indices, and
  are hence only present for $M>1$ bands:
  density-density interaction $V_0={\cal
    V}_{iiii}^{\alpha\beta\alpha\beta}$,
  direct exchange interaction $F_0={\cal
    V}_{iiii}^{\alpha\beta\beta\alpha}$, and
  hopping of double occupancies $F_0'={\cal
    V}_{iiii}^{\alpha\alpha\beta\beta}$.  For
  simplicity we assume the orbitals to be
  equivalent, i.e.\ $U$ is the same for all
  orbitals $\alpha$, and $V_0$, $F_0$, $F_0'$
  each have a fixed value for all pairs of
  orbitals $\alpha$, $\beta$.  Furthermore it
  should be noted that for equivalent orbitals
  these parameters are not independent, but
  are related by $U=V_0+2F_0$, $F_0=F_0'$
  \cite{Oles83}. In addition to these Hund's
  rule couplings there are still the
  inter-site terms, namely the hopping
  $t^{\alpha}_{ij}$ which takes place only
  between like orbitals (this follows from the
  general derivation above), and the
  next-neighbor interactions $V_1={\cal
    V}_{ijij}^{\alpha\alpha\alpha\alpha}$,
  $X_1={\cal
    V}_{iiij}^{\alpha\alpha\alpha\alpha}$,
  $F_1={\cal
    V}_{ijji}^{\alpha\alpha\alpha\alpha}$,
  $F_1'={\cal
    V}_{iijj}^{\alpha\alpha\alpha\alpha}$. For
  simplicity, these next-neighbor couplings
  are assumed to be diagonal in the band
  indices, i.e.\ they act only between like
  orbitals on neighboring sites.  Finally,
  since we are dealing with equivalent
  orbitals, we assume that the next-neighbor
  parameters $t$, $X_1$, $F_1$, $F_1'$ each
  have the same value for all bands $\alpha$.
  
  The resulting multi-band Hamiltonian then reads
  \begin{equation}
    \label{mbham}
    \hat{H}_{\mbox{\scriptsize\em M-band}}
           ^{\mbox{\scriptsize\em NN}}=
    \sum_{\alpha=1}^{M}
    \hat{H}^{\mbox{\scriptsize\em NN}}
    _{\mbox{\scriptsize\em 1-band,$\alpha$}}
    +\hat{H}_{\mbox{\scriptsize\em interband}}
  \end{equation}
  where
  \begin{eqnarray}
    \label{alphn}
    \hat{H}^{\mbox{\scriptsize\em NN}}
    _{\mbox{\scriptsize\em 1-band,$\alpha$}}
    &=&
    -t    \sum_{\langle i,j\rangle,\sigma}
          (\hat{c}^+_{i\alpha\sigma}
          \hat{c}^{\phantom +}_{j\alpha\sigma}
          +\mbox{h.c.})
    \nonumber\\
    &+&U  \sum_{i\sigma}
          \hat{n}_{i\alpha\downarrow}
          \hat{n}_{i\alpha\uparrow}
    +\sum_{\langle i,j\rangle}
    \Big[
    V_1   \hat{n}_{i\alpha}\hat{n}_{j\alpha}
    \nonumber\\
    &+&X_1  \sum_\sigma(\hat{c}^+_{i\alpha\sigma}
          \hat{c}^{\phantom +}_{j\alpha\sigma}
          +\mbox{h.c.})
          (\hat{n}_{i\alpha-\sigma}+\hat{n}_{j\alpha-\sigma})
    \nonumber\\
    &+&F_1'(\hat{c}^+_{i\alpha\uparrow}
          \hat{c}^+_{i\alpha\downarrow}
          \hat{c}^{\phantom +}_{j\alpha\downarrow}
          \hat{c}^{\phantom +}_{j\alpha\uparrow}
          +\mbox{h.c.})
    \nonumber\\
    &-&2F_1 (\hat{\bf S}_{i\alpha}\hat{\bf S}_{j\alpha}
             +\frac{1}{4}\hat{n}_{i\alpha}\hat{n}_{j\alpha})
   \Big]
  \end{eqnarray}
  is a straightforward generalization of the
  one-band Ha\-mil\-tonian (\ref{nn}) to more
  than one band, and
  \begin{eqnarray}
    \label{atham}
    \hat{H}_{\mbox{\scriptsize\em interband}}&=&
    \sum_{i;\alpha<\beta}\Big[
      V_0   \hat{n}_{i\alpha}\hat{n}_{i\beta}
      \nonumber\\
      &-&2F_0 (\hat{\bf S}_{i\alpha}\hat{\bf S}_{i\beta}
             +\frac{1}{4}\hat{n}_{i\alpha}\hat{n}_{i\beta})
             \nonumber\\
      &+&F_0' (\hat{c}^+_{i\alpha\uparrow}
             \hat{c}^+_{i\alpha\downarrow}
             \hat{c}^{\phantom +}_{i\beta\downarrow}
             \hat{c}^{\phantom +}_{i\beta\uparrow}
             +\mbox{h.c.})\Big]\,.
  \end{eqnarray}
  Several of the processes contained in the
  Hamiltonian (\ref{mbham}) are illustrated in
  Fig.\ \ref{orb}.
  
  The physical picture of bulk ferromagnetism, originally put forward
  by Slater, is the following. If the on-site Hund's rule couplings
  are strong enough, they lead to an independent ferromagnetic
  alignment of spins on each atom. In this situation the kinetic
  energy plays a decisive role since it can serve to communicate the
  spin alignment across the solid. Indeed, if the alignment on
  neighboring atoms were different, the hopping of electrons from one
  atom to the next would generate on-site Hund's rule interactions and
  thus increase the energy. These interactions can only be avoided if
  the spin alignment on neighboring atoms is the same, implying {\em
    global} ferromagnetism. This mechanism for the stabilization of
  ferromagnetic order works all the better the larger the number $M$
  of orbitals, i.\ e. bands, is onto which an electron can hop; it
  does not work for a single band ($M=1$).  Therefore one should
  expect band degeneracy to favor ferromagnetism.  This very
  qualitative picture is indeed found in Stoner mean-field theory
  \cite{Cyrot73}: Since the non-interacting DOS is proportional to the
  number of degenerate bands, the critical interaction decreases as
  $U_c^{\mbox{\scriptsize\em Stoner}}(M)\sim 1/M$.  On the other hand,
  few {\em exact} results are known in the case of degenerate bands.
  For example, for a one-dimensional chain with two orbitals ($M=2$)
  and infinite on-site Coulomb interactions it has been shown that the
  ground state is ferromagnetic for $N=L+1$ electrons
  \cite{Lacroix76}. This statement has been extended to $L+1\leq
  N\leq2L-1$, i.e.\ up to one electron less than half-filling
  \cite{Kubo82}. The important point is that on a one-dimensional
  chain with only one orbital and hopping between nearest neighbors,
  Nagaoka's theorem is not applicable, since the lattice does not have
  loops. In that case the ground state is degenerate with respect to
  the total spin $S$.
  On the other hand, if there are two orbitals the loop property is
  restored, and the ferromagnetic states with maximum spin become the
  only ground states.
  
  Furthermore, the following rigorous result can be
   established for the Hamiltonian (\ref{mbham}) at
   half-filling \cite{Kollar97etc}. For $N=ML$ electrons, the
   ground states of $\hat{H}_{\mbox{\scriptsize\em
       M-band}}$ have maximum spin
   $S=S_{\mbox{\scriptsize\em max}}=ML/2$ if
   \begin{equation}
     2V_0\geq F_0\geq\frac{U_c^{(1,2)}}{1+M/2}
     \;\;\;\mbox{and}\;\;\;F_1>0
   \end{equation}
   where $U_c^{(1,2)}$ are the critical values for $U$ in
   the {\em single} band system, as given in
   (\ref{bounds1})-(\ref{bounds2}). The meaning of these
   bounds is the following. The requirement that
   $2V_0\geq F_0$ and $F_0$ be greater than a certain
   threshold leads to an alignment of the electronic
   spins on an isolated atom. On the other hand,
   ferromagnetism within each band is brought about by
   the next-neighbor exchange $F_1>0$ and the Hubbard
   interaction $U$ larger than a threshold related to
   $U_c^{(1,2)}$. The combination of these two effects
   (using the fact that $U=V_0+2F_0$ for equivalent bands)
   leads to a critical value for the Hund's rule
   coupling $F_0$, which indeed becomes lower as the
   number of bands increases.

   While this result contains some ideas of
   Slater's picture, it does not explain the
   itinerant aspects of multi-band ferromagnetism, since at
   half-filling the ferromagnetic ground states are
   insulating. So far, this result can only be modified to
   apply also to Nagaoka's case (one hole, $N=ML-1$)
   \cite{Kollar97etc}.

 \subsection{Kinetic energy and lattice structure}

The stability of ferromagnetism is intimately linked with the structure of the 
underlying lattice and the kinetic energy (i.e.~the hopping) of the electrons
\cite{Hirsch94etc,Hlubina96etc,Daul97etc,Shastry90,Fazekas90,MuellerHartmann93,Nagaoka66,Tasaki95,Penc96,Fazekas96etc}. This is supported by several facts: (a) Nagaoka's proof of 
ferromagnetism
in the Hubbard model for a single hole at $U=\infty$ \cite{Nagaoka66} 
depends on the
existence of closed loops along which the hole can move, and (b) on bipartite 
lattices $anti$\-ferro\-magnetism is the generic magnetic state making it hard
for ferromagnetism to become stable. Hence $non$-bipartite lattices with loops
(or with a kinetic energy involving hopping between nearest $and$ next-nearest 
neighbors sites effectively leading to a motion on loops) should be expected
to support ferromagnetism because the competing anti\-ferro\-magnetic 
tendencies are severely weakened, and because the corresponding DOS of 
non-interacting electrons is asymmetric and thus has a peak at a non-symmetric
position. Indeed, a peak at one of the band edges as in the case
of the fcc lattice is favorable for ferromagnetism
\cite{Gutzwiller63etc,Macedo92,Hirsch94etc,Shastry90,Fazekas90,MuellerHartmann93}.
This is supported by the observation \cite{MuellerHartmann93} that Co and Ni,
having non-bipartite hcp and fcc lattice structure, respectively, show a full
magnetization while bcc-Fe has only a partial magnetization.

\subsubsection{A model density of states}

To gain insight into why a DOS with a peak (or more precisely with a high
spectral density) at the band edge,
e.g. at the lower band edge for $n<1$, may be favorable for 
ferromagnetism, we study non-interacting electrons with the following model DOS:
\begin{equation}
 N^0(E) = \frac{1}{\Delta} \left(1+\frac{1}{3}A^2+\frac{2A}{\Delta}E\right)
\end{equation}
where $\Delta$ is the width of the band and $A$ parameterizes the asymmetry.
The lower band edge is $-(A+3)\Delta/6$ such that the first moment of
$N^0(E)$ vanishes.
For $A=-1 (+1)$ the DOS has a triangular shape with the peak at the lower 
(upper) edge, while for $A=0$ it is flat.
We wish to calculate the energy difference $\delta \epsilon$ between the fully
polarized and the paramagnetic state
\begin{equation}
 \delta \epsilon \equiv \epsilon_{\mbox{\scriptsize\em ferro}}(n,A)-
 \epsilon_{\mbox{\scriptsize\em para}}(n,A)
\end{equation}
as a function of $A$ and the band filling.
It is easy to confirm that for all $n$ $\delta \epsilon$ is lowest for 
$A=-1$.
The reason is this:
In the paramagnetic state $N$ non-interacting electrons 
($N/2$ electrons with spin up and down each) fill the lowest $N/2$ 
$\vec k$-states up to an energy $E_F^{\mbox{\scriptsize\em para}}$, 
while in the ferromagnetic state the $N$ singly occupied $\vec k$-states below 
$E_F^{\mbox{\scriptsize\em ferro}}>E_F^{\mbox{\scriptsize\em para}}$ are 
filled, i.e.~the $N/2$ $\vec k$-states with energy above 
$E_F^{\mbox{\scriptsize\em para}}$ are also occupied.
The higher the DOS is at the lower edge the less the additional 
$N/2$ $\vec k$-states are forced into high-energy states, 
i.e.~the lower $E_F^{\mbox{\scriptsize\em ferro}}$
will be. Then $\delta \epsilon$ is kept at a minimum.

 \subsubsection{The Hubbard model on fcc type lattices}

 As explained in the previous section, a lattice structure which gives
 a high DOS at low energies is expected to be favorable
 for ferromagnetism (at $n<1$) because this situation reduces 
 the loss in kinetic energy.
 The question remains: is a strongly peaked DOS \em sufficient \em to induce
 ferromagnetic order in the single band Hubbard model without additional
 interactions? 

 In this section we discuss the results of a QMC-investi\-gation 
 of the stability
 of ferromagnetism in the single band Hubbard model in the limit of
 infinite dimensions. 

Based on the considerations in Sec.~3.3.1, the fcc lattice is expected to be
a good candidate
for ferromagnetism because of its high (divergent) DOS at the lower band edge. 
The fcc-lattice can be generalized to higher dimensions in different ways
\cite{MuellerHartmann91etc,Uhrig96}. 
Here we use the definition of an fcc lattice
 as the set of all points with integer cubic coordinates summing up to an even
 integer \cite{MuellerHartmann91etc}. 
 It is a non-bipartite Bravais lattice for any dimension $d>2$.
 For $d=2$ it is identical to the square lattice. 
 Nearest neighbors are connected by \em two different \em unit vectors on
 a simple hypercubic (hc) lattice. The coordination number is hence
 $Z=2d(d-1)$.
 With the proper scaling of the hopping term, (\ref{scale}),
 the non-interacting DOS of the generalized fcc lattice 
 can be calculated in $d=\infty$ \cite{MuellerHartmann91etc} as:
 \begin{equation}
 N^0_{\mbox{\scriptsize\em gfcc}}(E) = e^{-(1+\sqrt{2} E)/2}/\sqrt{\pi (1+\sqrt{2} E)} 
 \label{glferro1}
 \end{equation}
 which has a strong square-root divergency at the lower band edge, 
 $-1/\sqrt{2}$, and no upper band edge.
 Since we choose a \em positive \em hopping integral $(-t>0)$, $N^0_{\mbox{\scriptsize\em gfcc}}(E)$ 
 might be regarded as the DOS of holes rather than electrons. 
 A full polarization of holes would hence correspond to  
 a \em maximal \em (though not full) polarization in a more than 
 half-filled band.

 While the three dimensional fcc lattice has no square-root but only
 a logarithmic divergency at the band edge it is worth mentioning that 
 a square-root divergency arises on the fcc lattice in any dimension 
 if there is an
 additional next nearest neighbor hopping of the size $t'=t/2$ between 
 sites that are linked by two unit vectors in the \em same \em direction 
 on the hc lattice.
 The energy dispersion and the DOS of this so-called ``half-hypercubic'' (hh) 
 lattice \cite{Uhrig96} can easily be obtained from the hc lattice as:
 \begin{eqnarray}
 \epsilon_{hh}(k) & = & \frac{t}{2t_{hc}} \epsilon_{hc}^2(k) - 3t 
 \label{glferro2} \\
 N^0_{hh}(E)  
 & = & \frac{2t_{hc}^2}{t} \frac{N^0_{hc}(\sqrt{E+3t})}{\sqrt{E+3t}} \; . 
 \label{glferro3}
 \end{eqnarray}
 In the limit $d\to\infty$, the hh and fcc lattices become equivalent.

 The fact that the fcc lattice provides a good `environment' for
 ferromagnetism has also been supported by variational studies of the
 stability of the Nagaoka state \cite{Shastry90,MuellerHartmann93}.
 Variational calculations provide
 limits for the critical density $n_c$ and  
 the critical interaction $U_c$ where saturated ferromagnetism becomes 
 unstable:
 While for the hc lattice in $d=\infty$ the stability regime
 shrinks to the point $U_c=\infty$, $\delta_c=0$ \cite{Fazekas90}, 
 for the fcc DOS (\ref{glferro1}) there is a 
 critical line $U_c(n)$ with $U_c(0)=0$ and 
 $U_c(1)=\infty$ \cite{MuellerHartmann95etc}. 
 The Nagaoka state is always unstable in the case of electron doping $(n>1)$.

 Recently, these variational boundaries were qualitatively confirmed
 by Uhrig's calculation of the exact single spin-flip energy of the Nagaoka
 state in $d=\infty$ \cite{Uhrig96}. While on the hh lattice 
 $U_c$ vanishes at low densities, $U_c$ remains finite for all densities in the
 case of the ``laminated'' lattice which is a different generalization of the
 fcc lattice without a divergent DOS at the band edge.
 His results, too, emphasize the subtle dependence of the stability
 of ferromagnetism on the lattice structure. 

 Antiferromagnetism is not expected on the fcc-lattice in high dimensions
 because the difference of the numbers of not frustrated bonds and frustrated 
 bonds is only of the order of $d$ resulting in an effective field of the 
 order of $t^2 d\propto 1/d$ \cite{MuellerHartmann95etc}.
 Even in $d=3$ antiferromagnetism is frustrated and is 
 expected only very close to half-filling.

 To detect a ferromagnetic instability we calculated the temperature dependence 
 of the uniform static 
 susceptibility, $\chi_F$, from the two-particle correlation functions
 \cite{Ulmke95a}. At an intermediate interaction strength of $U=4$ we found the 
 ferromagnetic response to be strongest around quarter filling ($n\simeq 0.5$).
 $\chi_F$ obeys a Curie-Weiss law (Fig.~\ref{figferro1}) and the Curie 
 temperature $T_c$ can safely be extrapolated from the zero of $\chi_F^{-1}$
 to a value of $T_c=0.051(2)$ at $n=0.58$.
 Below $T_c$ the magnetization $m$ grows rapidly, reaching more than
 80\% of the fully polarized value ($m_{max}=n=0.58$) at the lowest
 temperature which is only 30\% below $T_c$.
 The three data points $m(T)$ (Fig.~\ref{figferro1}) are consistent with a 
 Brillouin function with the same critical temperature 
 of $T_c=0.05$ and an extrapolated full polarization at $T=0$.
 A saturated ground state magnetization is also consistent with 
 the single spin-flip energy
 of the fully polarized state which is positive at the present parameter
 values~\cite{Uhrig96}.

 Translated to the three dimensional fcc lattice with $Z=12$ nearest
 neighbors and a bandwidth of $W=16t$, the critical temperature becomes
 $T_c(3D)\approx 0.011 W$~\cite{U4comment}.
 Thus, despite the oversimplifications of the single band Hubbard model,
 the resulting Curie temperature has a realistic order of magnitude of 
 500-800K for typical values of $W$ around 5 eV. 

 In order to answer the question if the system is still metallic
 we also calculated the single particle spectrum \cite{Ulmke96etc}.
 We find that the system is metallic since both spectra have a finite value
 at the Fermi level. 

 We conclude that within the dynamical mean-field theory 
 a DOS with sufficiently large spectral weight at low energies is able
 to induce itinerant ferromagnetism in the single band Hubbard model.
 Detailed investigations of the dependence of ferromagnetism on the 
 electronic density, the interaction strength, and in particular 
 the extension to more realistic densities of states 
 (e.g.~fcc lattice in $d=3$) is under progress and will be presented elsewhere
 \cite{Ulmke96etc}.

 In the $d=\infty$ limit the dynamics of the correlated system is fully
 taken into account, while spatial fluctuations are suppressed
 \cite{MuellerHartmann89etc,Janis91,Georges92a,Jarrell92,Janis92a}.
 One might therefore suspect that the stability of ferromagnetism 
 is somehow $over$\-estimated in this approach (in particular since 
 the rivaling anti\-ferro\-magnetism is completely absent on the fcc lattice).
 However, most recently similar results were reported for the Hubbard model
 with nearest and next-nearest neighbor hopping $t$ and $t'$, respectively,
 in dimensions $d=1$ \cite{Daul97etc} and $d=2$ \cite{Hlubina96etc},
 which are consistent with the results in $d=\infty$. Hence the existence of 
 itinerant ferromagnetism in the pure Hubbard model with a suitable kinetic
 energy seems to be established at last.

 \section{Metamagnetic phase transitions}

An issue related to the stability of (itinerant) ferromagnetism in the Hubbard
model is the question concerning metamagnetism \cite{Becquerel39} 
in this and other models. Here it is
an external magnetic field $H$ which helps to suppress the (not necessary 
long-range) antiferromagnetic correlations in the system and thereby induces 
a pronounced transition from a state with low magnetization to one with high 
magnetization. At the metamagnetic transition the magnetization curve $m(H)$ 
shows an up-turn such that the susceptibility $\chi(H)=\partial m / \partial H$
has some kind of maximum. This feature serves as a 
convenient general definition for ``metamagnetism''. Metamagnetic transitions 
were first observed in 
{ strongly anisotropic antiferromagnets}  of which
FeCl$_2$ and Dy$_3$Al$_5$O$_{12}$ (DAG) are well-studied
prototypes \cite{Stryjewski77etc}.
These materials are insulators where the valence electrons
are localized at the Fe and Dy ions, respectively.
 The arising local moments order {antiferromagnetically} and are 
 { strongly anisotropic} in the sense that they are constrained to lie
 along an easy axis ${\mathbf e}$. In this case a spin-flop transition 
 in an external magnetic field  ${\mathbf H} \parallel {\mathbf e}$
 cannot occur. 
 Apart from the above materials there are also conducting systems
 that most probably belong to
 this class, e.g.~the conductors UA$_{1-x}$B$_x$
 (where $\mathrm A =P$, $\mathrm As$; $\mathrm B = S$, 
 $\mathrm Se$) \cite{Stryjewski77etc}, SmMn$_2$Ge$_2$ \cite{Brabers94} 
 and  TbRh$_{2-x}$Ir$_x$Si$_2$ \cite{Ivanov95}.

 Hitherto completely different theories are employed
 to describe metamagnetic phase transitions in these
 different, strongly anisotropic antiferromagnets.
 Investigations of localized systems
 are usually based on the
 Ising model, where more than one interaction
 has to be introduced to describe the experimentally observed
 first order phase transitions \cite{Lawrie84etc}.
 With  antiferromagnetic
  coupling $J$ between the $Z$ nearest-neighbor (NN) spins 
 and a ferromagnetic 
 coupling $J'$ between the $Z'$ next-nearest-neighbors (NNN)
 one obtains 
 \begin{equation}
 H_{\mbox{\scriptsize\em Ising}} = 
J \sum_{NN} S_iS_j - J' \sum_{NNN} S_i S_j - 2 H \sum_i S_i \; .
 \label{isi}
 \end{equation}
 In Weiss mean-field theory this model
 shows two different types of phase diagrams
 depending on the parameter $R \equiv Z'J'/(ZJ)$ \cite{Kincaid74etc}.
 For $R>3/5$ the first and second order phase transition
 line join smoothly at the same point, producing  
 a tricritical point (TCP),
 while for  $R<3/5$ there is no common endpoint
 (see Fig.~\ref{scheme}).
 However the scenario of  Fig.~\ref{scheme}b was not found
 when evaluating (\ref{isi}) beyond mean-field theory \cite{Herrmann93etc}.

 For {\em itinerant electron metamagnetism} (IEM) Moriya and Usami 
 \cite{Moriya77}
 proposed a Landau theory, where the parameters have to be deduced from
 microscopic models. The idea is to calculate first the independent
 electron band structure and to introduce the Coulomb interaction within the
 random phase approximation.

 It is our purpose to investigate the origin of
 metamagnetism in strongly anisotropic antiferromagnets from a
 microscopic, quantum-mechanical point of view, and to
 describe  different kinds of 
 metamagnets (i.e.~metallic {\em and} insulating,
 band-like {\em and} localized systems)
 qualitatively within a single model.
 To this end we study the Hubbard model (\ref{hub}) with the additional 
 constraint that the antiferromagnetic magnetization
 ${\mathbf m}_{st}$  lies {\it parallel} to the external magnetic field
 ${\mathbf H}$ \cite{Giesekus93etc}.
 In this way the existence of an { easy axis} ${\mathbf e}$ along which 
 ${\mathbf H}$ is directed, such that ${\mathbf e} \parallel
 {\mathbf m}_{st} \parallel {\mathbf H}$, is incorporated
 in a natural way. 
 By this approach, both kinetic energy and Coulomb interaction
 are captured microscopically, whereas the relativistic corrections
 (responsible for the easy axis) are not. 
 This procedure is justified since the relativistic corrections are
 of ${\cal O}(10^{-2}$ eV) and are thus small compared to the  
  kinetic and Coulomb energy which are of ${\cal O}(1$eV).
 Therefore the existence of an anisotropy axis
 ${\mathbf e}$ and the correlations
 described by the  Hubbard model  are quite unrelated.
 Note, that the existing Ising
 and IEM theories do not treat the  kinetic energy and Coulomb
 interaction microscopically, but within an effective model.

 A perturbative treatment of the Hubbard model with easy axis in the weak and
 strong coupling limit shows that the appearance of a tri- or multicritical 
 point is a delicate matter, since neither of these two limits is able to 
 describe a change of the transition from first to second order \cite{Held96}. 
 Apparently the entire 
 transition scenario depends sensitively on the value of
 the electronic on-site interaction $U$. To study this point in
 greater detail we have to go to intermediate coupling.
 In this non-perturbative regime we employ again QMC
 simulations to calculate the magnetization $m(H)$ and the staggered 
 magnetization $m_{st}(H)$ of the Hubbard model (\ref{hub}) in $d=\infty$ 
 \cite{Held96}.
 As the results do not much depend on the precise form
 of the density of states
 we choose $N^0(\epsilon) = [(2t^*)^2 - \epsilon^2 ]^{1/2}/(2 \pi t^{*2})$,
 setting $t^* \equiv 1$ in the following. 
 All calculations are performed at half-filling.

 The results for $m(H)$ and $m_{st}(H)$
  are used to construct the $H-T$ phase diagram at $U=4$ 
 (Fig.~\ref{hvst}). It displays all the features of Fig.~\ref{scheme}b.
 In particular, the first order line  continues {\it into}
 the ordered phase, separating two different AF phases: AF$_I$
 (where $ m \simeq 0$) and AF$_{II}$ (where $m > 0$).  The position 
 of its  endpoint  cannot, at
 present, be determined accurately (dotted line).
 This phase diagram is
 surprisingly similar to the experimental phase diagram of FeBr$_2$ 
 \cite{Azevedo95etc}. 
 
 A change in $U$ and the filling $n$ will affect the phase
 diagram quantitatively and qualitatively. 
 These results will be reported elsewhere \cite{Held97etc}.

 \section{Conclusion}
 In the last few years, and especially most recently, considerable
 progress was made in our understanding of the microscopic origin of itinerant 
 ferromagnetism.
 These results were obtained on the basis of well-defined lattice models of 
 correlated electrons, of which the one-band Hubbard model is a particularly 
 important ingredient, by applying new, non-perturbative techniques, ranging 
 from rigorous to large-scale numerical methods. 
 There exists convincing evidence now that on appropriate, 
 non-artificial lattices, or for an appropriate kinetic energy, itinerant 
 ferromagnetism is stable even in the pure Hubbard model, for electronic 
 densities not too close to half-filling and large enough $U$.
 Important ingredients are: \\
 (i) lattices with loops 
 (or a kinetic energy allowing for motion on loops, e.g.~with $t,t'$ hopping) 
 such that the Nagaoka mechanism works and 
 antiferromagnetism is suppressed,\\
 (ii) a large spectral weight near the band edge. (We note that this condition
 goes far beyond the mean-field Stoner criterion for ferromagnetism, $UN(0)=1,$
 where $N(0)$ is the DOS at the Fermi energy.)

 The direct exchange interaction, as well as band degeneracy, will strongly 
 reduce the critical value of $U$ above which ferromagnetism becomes stable.

 Furthermore, using the dynamical mean-field theory
 to solve the Hubbard model with easy-axis 
 in the intermediate coupling regime, it appears to be possible to describe 
 various,  different forms of metamagnetism within a 
 single microscopic theory. 

 The development and application of controlled, 
 non-perturbative techniques will continue to be of particular importance 
 for the investigation of correlated electron systems.

\section*{Acknowledgement}
 One of us (DV) thanks Elliott Lieb for very useful discussions. In its early 
 stages this work was supported in part by the Deutsche Forschungsgemeinschaft 
 under SFB 341.

\newpage

\begin{table}[h]
  \caption{\label{crtab}Sufficient conditions for 
    ferromagnetic ground states with one hole.}
\end{table} 

 \begin{table}[h]
     \begin{center}
       \begin{tabular}{lll}
         \hline
         1a&$\!\!F>0$, $X\neq t$, $U>U_c^{(1)}$
         \begin{tabular}{l}
           any $t$\\
           $t\leq0$
         \end{tabular}
         \begin{tabular}{l}
         bipartite lattice\\
         non-bip. lattice
       \end{tabular}
       \\ \hline
       1b&$\!\!F>0$, $X=t$, $U\geq U_c^{(2)}$
       \begin{tabular}{l}
         any $t$\\
         $t\leq0$
       \end{tabular}
       \begin{tabular}{l}
         bipartite lattice\\
         non-bip. lattice
       \end{tabular}
       \\ \hline
       2&$\!\!F=0$, $X=t$, $U>U_c^{(2)}$
       \begin{tabular}{l}
         $t\neq0$\\
         $t<0$
       \end{tabular}
       \begin{tabular}{l}
         bipartite, with loops\\
         non-bip., with loops
       \end{tabular}
       \\ \hline
     \end{tabular}
   \end{center}
   \end{table}

\bigskip

 \begin{figure}[h]
    \caption{\label{uvf}Critical value $U_c$ vs.\ exchange
      interaction $F$ for different $t$, $V$, $X$, and
      $F'=0$. For $U>U_c$ the ground state is ferromagnetic.
      See Table~\protect\ref{crtab} for details.}

\medskip

\caption{Critical value of the direct exchange coupling $F^*$ above which 
ferromagnetism is stable vs. Hubbard interaction $U$. The phase boundaries 
were calculated from the divergence of the ferromagnetic (diamonds) and 
antiferromagnetic (squares) susceptibilities, respectively. The QMC 
results were
extrapolated to $T=0$, the filling is $n=1$ (F: ferromagnetic phase, 
AF: antiferromagnetic phase, P: paramagnetic phase).
Dotted line: Hartree(-Fock) theory; dashed line: Heisenberg limit.
\label{phaseT0}
}

\medskip

    \caption{\label{orb}Illustration of the interaction
      and kinetic energy in a two-band model: $U$,
      $V_0$, $F_0$ are the Hubbard interaction, the
      density-density interaction, and the direct
      exchange interaction, respectively, acting on
      the same site, while $V_1$ and $F_1$ act
      between neighboring sites $i$ and $j$, and
      $t$ is the hopping.}

\medskip

\caption{
Magnetization  $m$ (diamonds) and inverse ferromagnetic
susceptibility  $\chi_F^{-1}$ (circles; values multiplied by a factor of 4
to use the same scale) for $U=4$ and $n=0.58$. Error-bars are of the size of
the symbols or smaller. (Note that the value of $\chi_F^{-1}$ at $T=0.05$ is 
a data point, not an extrapolation.)
The dotted line is a linear fit to $\chi_F^{-1}$,
the dashed line a fit with a Brillouin function to $m$.
\label{figferro1}
}

\medskip

\caption{Schematic $H-T $ phase diagram for
a) a typical Ising-type metamagnet (TCP: tricritical point), 
b) the Ising model ({\protect\ref{isi}}) in
mean-field theory with $R < 3/5$ (CE: critical endpoint, BCE:
bicritical endpoint) \protect\cite{Kincaid74etc}. 
Full lines:  first order transition, 
broken lines: second order
transition; AF: antiferromagnetic phase, P: paramagnetic phase.
\label{scheme}
}

\medskip

\caption{
$H-T$ phase diagram for
the $d = \infty$ Hubbard model with easy axis along $H$ at $n = 1$ 
and $U = 4$  \protect\cite{Held96}; same notation as in Fig. 
\protect\ref{scheme}b.
\label{hvst}}

\end{figure}

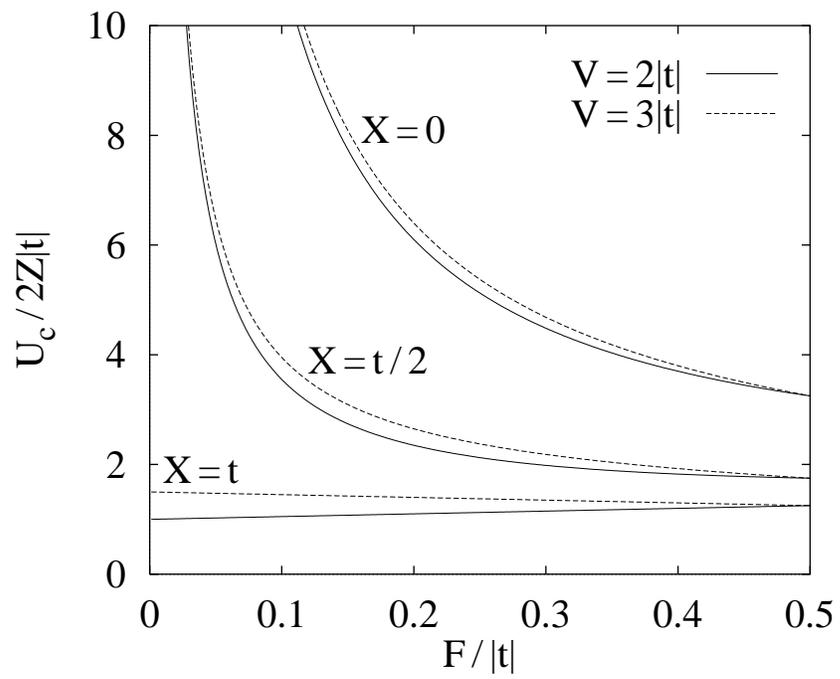
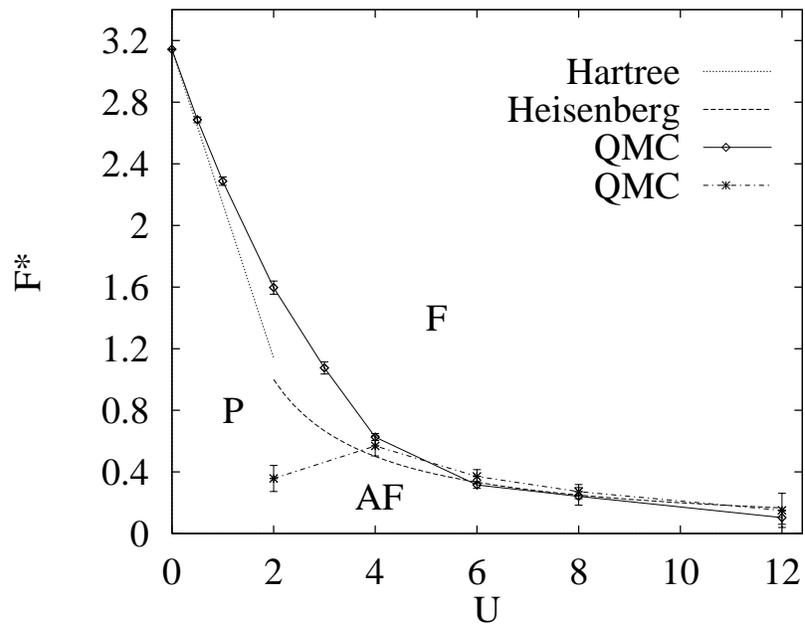
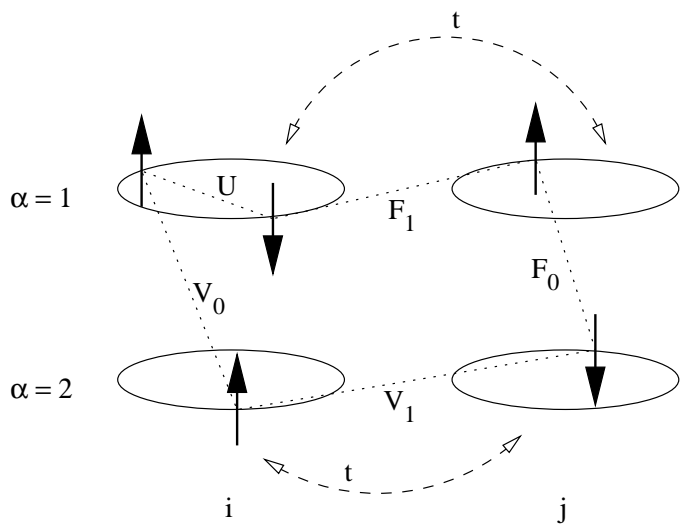
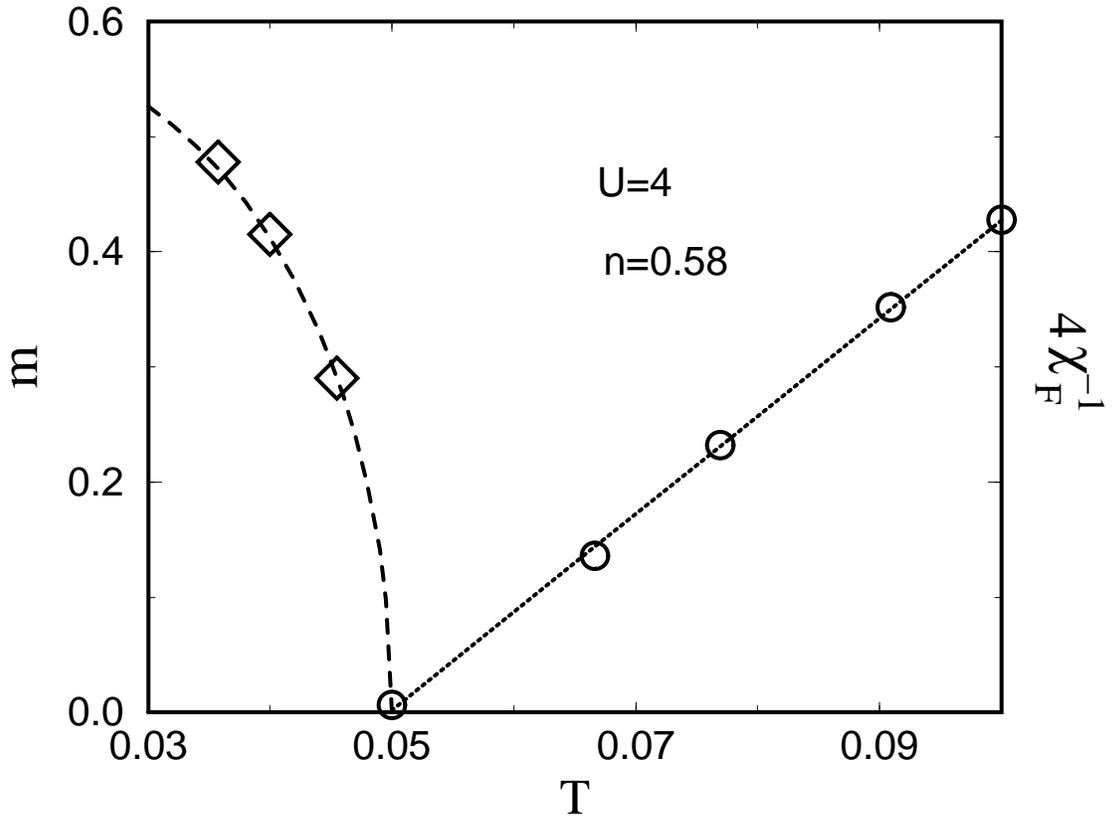
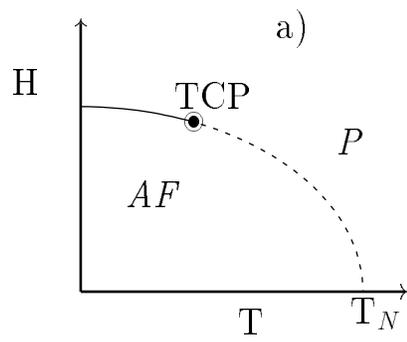
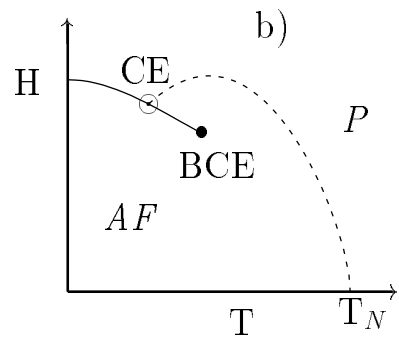
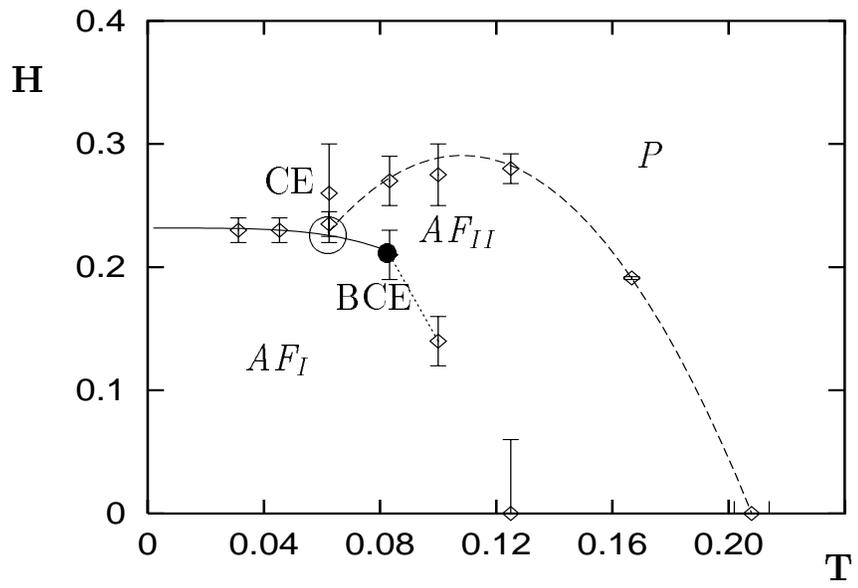
\end{document}